# Periodic orbits around areostationary points in the Martian gravity field

Xiaodong Liu[1], Hexi Baoyin[2], and Xingrui Ma[3]

School of Aerospace, Tsinghua University, 100084 Beijing, China

Email: liu-xd08@mails.tsinghua.edu.cn; baoyin@tsinghua.edu.cn;

maxr@spacechina.com

## Abstract

This study investigates the problem of areostationary orbits around Mars in the three-dimensional space. Areostationary orbits are expected to be used to establish a future tele-communication network for the exploration of Mars. However, no artificial satellites have been placed in these orbits thus far. In this paper, the characteristics of the Martian gravity field are presented, and areostationary points and their linear stability are calculated. By taking linearized solutions in the planar case as the initial guesses and utilizing the Levenberg-Marquardt method, families of periodic orbits around areostationary points are shown to exist. Short-period orbits and long-period orbits are found around linearly stable areostationary points, and only short-period orbits are found around unstable areostationary points. Vertical periodic orbits around both linearly stable and unstable areostationary points are also examined. Satellites in these periodic orbits could depart from areostationary points by a few degrees in longitude, which would facilitate observation of the Martian topography. Based on the eigenvalues of the monodromy matrix, the evolution of the stability index of periodic orbits is determined. Finally, heteroclinic orbits connecting the two unstable areostationary points are found, providing the possibility for orbital transfer with minimal energy consumption.

---

[1] PhD candidate, School of Aerospace, Tsinghua University
[2] Associate Professor, School of Aerospace, Tsinghua University
[3] Professor, School of Aerospace, Tsinghua University

## 1. Introduction

Stationary orbits are approximately equatorial and circular, and share the same period with the planetary rotation. Satellites in stationary orbits remain in orbit over the same location on the planetary surface. The notion of stationary orbits for Earth was first proposed by Clarke, and these orbits were considered to be a good application for communication satellites (Clarke 1945). Earth's geostationary orbits have been widely used in practical engineering applications, including communication, navigation, and meteorological applications. Geostationary orbits have been studied for many years, and plenty of papers contributed to this research.

Stationary orbits around Mars are also known as areostationary orbits, and their characteristics are similar to Earth's geostationary orbits. Areostationary orbits are expected to be used to establish a future tele-communication network for the exploration of Mars (Edwards et al. 2000; Bell et al. 2000; Hastrup et al. 2003; Turner 2006; Edwards & Depaula 2007; Edwards 2007). However, no artificial satellites have been placed in these orbits thus far, and not many studies concerning stationary orbits around Mars have been conducted.

The positions of stationary points of Earth, Moon, and Mars were found to be singular points of the zero-velocity surface (Zhuravlev 1977). For Earth and Mars, Deprit and López-Moratalla proved that the two linearly stable points are also stable in the sense of Liapunov (Deprit & López-Moratalla 1996). The influences of the individual harmonics of the Martian potential on the positions of stationary points were examined qualitatively and quantitatively, in which the $C_{20}$, $C_{22}$, and $S_{22}$ terms were considered in the planar case (Wytrzyszczak 1998). In other research on areostationary orbits, the drift in longitude due to spherical harmonics up to the $2^{nd}$ degree and order was analyzed, and both the period of libration and the amount of stationkeeping maneuvers were calculated (Alvarellos 2008). Liu et al. also analyzed the longitudinal drift in the Martian $2^{nd}$ degree and order gravity field, and more accurate stationary points for reducing drift in the full gravity field were obtained using numerical methods (Liu et al. 2010). In addition, periodic orbits around equilibrium points in the Earth $2^{nd}$ degree and order gravity field (Lara & Elipe 2002) and in arbitrary $2^{nd}$ degree and order uniformly rotating gravity field (Hu & Scheeres 2008) have been studied for the planar motion.

This paper investigates areostationary orbits in the three-dimensional space. The major concerned perturbation is the effect of the Martian gravity field, and the ways in which the gravitational potential of Mars independently influences areostationary orbits are explored. The results can be used as a starting point for correction methods when considering all the perturbations including three-body perturbations, atmospheric forces, and solar radiation pressure. By taking linearized solutions in the planar case as the initial guesses and utilizing the Levenberg-Marquardt method (Levenberg 1944; Marquardt 1963; Moré 1977), it is demonstrated that periodic orbits exist around both unstable and linearly stable areostationary points, and their stability index are also calculated. Finally, heteroclinic orbits connecting the two unstable areostationary points are found, providing the possibility for orbital transfer with minimal fuel expenditure.

## 2. The Martian gravity field

The gravitational potential $U$ in the body-fixed reference frame is expressed as follows (Kaula 1966):

$$U = \frac{\mu}{r}\left[1 + \sum_{l=1}^{\infty}\sum_{m=0}^{l}\left(\frac{R_e}{r}\right)^l P_{lm}(\sin\varphi)(C_{lm}\cos m\lambda + S_{lm}\sin m\lambda)\right], \qquad (1)$$

where $\mu$ is the Martian gravitational constant; $R_e$ is the reference radius of Mars; $r$ is the position of the spacecraft; $P_{lm}$ is the associated Legendre function of degree $l$ and order $m$; $C_{lm}$ and $S_{lm}$ are the coefficients of the spherical harmonic expansion; $\varphi$ is the latitude of the body-fixed coordinate system; and $\lambda$ is the longitude of the body-fixed coordinate system. For Mars, the prime meridian is defined as the longitude of the crater Airy-0.

The recently improved spherical harmonic Martian gravity field used in this paper is MRO110B2 (Konopliv et al. 2011). The values of some first coefficients of MRO110B2 are given in Table 1. The gravity coefficients are normalized and are related to the unnormalized coefficients as follows (Kaula, 1966):

$$\begin{pmatrix} C_{lm} \\ S_{lm} \end{pmatrix} = \left[ \frac{(l-m)!(2l+1)(2-\delta_{0m})}{(l+m)!} \right]^{1/2} \begin{pmatrix} \overline{C}_{lm} \\ \overline{S}_{lm} \end{pmatrix}, \qquad (2)$$

where $\delta_{0m}$ is the Kronecker delta

$$\delta_{0m} = \begin{cases} 1, & \text{if } m = 0, \\ 0, & \text{if } m \neq 0. \end{cases} \qquad (3)$$

The value of the reference radius $R_e = 3396000$ m, and the value of the gravitational constant $\mu = 4.2828374527\text{E}+13$ m$^3$ s$^{-2}$ (Konopliv et al. 2011).

It is well known that the gravity field of Mars is farther away from the potential of a sphere than that of Earth (Wytrzyszczak 1998); therefore, areostationary orbits are more perturbed than geostationary orbits. According to the Martian gravity field MRO110B2, it is obvious that the Martian oblateness term $C_{20}$ is dominant among all harmonic coefficients, but it is not as dominant as Earth's $C_{20}$. The Martian $C_{30}$ term has the opposite sign and is stronger compared to Earth's $C_{30}$. Moreover, the Martian tesseral harmonics are also stronger than those of Earth. Unlike Earth, the Martian $C_{22}$ and $S_{22}$ terms have different signs.

Previous research has determined that the terms $C_{20}$, $C_{22}$, $S_{22}$, $C_{30}$, $C_{32}$, and $S_{32}$ are essential for areostationary orbits (Wytrzyszczak, 1998). The zonal harmonic $C_{20}$ causes the main shift of the position of areostationary points in the radial direction;

the tesseral harmonics $C_{22}$ and $S_{22}$ mainly influence the longitudinal distribution of areostationary points and the distances of areostationary points from Mars; the terms $C_{30}$, $C_{32}$, and $S_{32}$ contribute significantly to the displacement in the $z$ direction; and other harmonics only result in invisible changes in the position of areostationary points. The gravitational potential $U$ including these essential terms is written as follows:

$$U = \frac{\mu}{r}\left[1 + \frac{C_{20}R_e^2}{2r^2}\left(3\sin^2\varphi - 1\right) + \frac{3R_e^2}{r^2}\cos^2\varphi\left(C_{22}\cos 2\lambda + S_{22}\sin 2\lambda\right) \right. \\ \left. + \frac{C_{30}R_e^3}{2r^3}\left(5\sin^3\varphi - 3\sin\varphi\right) + \frac{15R_e^3}{r^3}\sin\varphi\cos^2\varphi\left(C_{32}\cos 3\lambda + S_{32}\sin 3\lambda\right)\right]. \tag{4}$$

**Table 1** Normalized values of certain normalized zonals and tesserals of the Martian gravity model MRO110B2.

| $l$ | $m$ | $\overline{C}_{lm}$ | $\overline{S}_{lm}$ |
| --- | --- | --- | --- |
| 2 | 0 | -0.8750219729111999E-03 | 0 |
| 2 | 2 | -0.8463591454722000E-04 | 0.4893448966831000E-04 |
| 3 | 0 | -0.1189641481101000E-04 | 0 |
| 3 | 2 | -0.1594791937546000E-04 | 0.8361425579193003E-05 |

## 3. Equations of motion

The motion of a satellite can be described in a rotating reference frame that is rotating with Mars. Accordingly, the rotating reference frame $Oxyz$ is established with the origin $O$ located at the center of mass of Mars, the $z$-axis taken as the rotation axis, and the $x$-axis coinciding with the prime meridian. It is assumed that Mars rotates uniformly around the $z$-axis with constant angular velocity $\omega$. Eq. (4) can be rewritten in the form of rectangular Cartesian coordinates as follows:

$$U = \frac{\mu}{r} - \frac{\mu R_e^2 C_2 \left(x^2 + y^2 - 2z^2\right)}{2r^5} + \frac{3\mu R_e^2 C_{22} \left(x^2 - y^2\right)}{r^5} \\ + \frac{6\mu R_e^2 S_{22} xy}{r^5} + \frac{15\mu R_e^3 C_{32} \left(x^2 - y^2\right)z}{r^7} + \frac{30\mu R_e^3 S_{32} xyz}{r^7}. \tag{5}$$

In mechanics, the effective potential $W$ is defined as the combined gravitational potential and rotational potential terms as follows (Scheeres et al. 2002; Thomas, 1993):

$$W = \frac{1}{2}\omega^2 \left(x^2 + y^2\right) + U. \tag{6}$$

Then, the Lagrangian $L$ of the motion is expressed as

$$L = T + W = \frac{1}{2}\left(\dot{x}^2 + \dot{y}^2 + \dot{z}^2\right) + \omega\left(x\dot{y} - \dot{x}y\right) + \frac{1}{2}\omega^2\left(x^2 + y^2\right) + U, \tag{7}$$

where $T$ is the kinetic energy. Scaling is performed so that $l = \left(\mu/\omega^2\right)^{1/3}$ is the unit of length and $1/\omega$ is the unit of time. With scaling, the Lagrangian $L$ is rewritten as

$$L = \frac{1}{2}\left(\dot{x}^2 + \dot{y}^2 + \dot{z}^2\right) + \left(x\dot{y} - \dot{x}y\right) + \frac{1}{2}\left(x^2 + y^2\right) + V, \tag{8}$$

where $V = U/\left(\omega^2 l^2\right)$. It is obvious that the Lagrangian is time-invariant, so the dynamical system admits the Jacobian integral $J$ as follows:

$$J = \frac{1}{2}\left(\dot{x}^2 + \dot{y}^2 + \dot{z}^2\right) - \frac{1}{2}\left(x^2 + y^2\right) - V. \tag{9}$$

The Lagrange equations are written as

$$\frac{\mathrm{d}}{\mathrm{d}t}\left(\frac{\partial L}{\partial \dot{\mathbf{r}}}\right) - \frac{\partial L}{\partial \mathbf{r}} = 0. \tag{10}$$

Therefore, the equations of motion of the spacecraft in the rotating reference frame

*Oxyz* can be written as

$$\ddot{x} - 2\dot{y} = x + V_x,$$
$$\ddot{y} + 2\dot{x} = y + V_y, \quad (11)$$
$$\ddot{z} = V_z.$$

## 4. Areostationary points

Areostationary points can be located by setting the right-hand sides of Eq. (11) to zero, i.e.,

$$x + V_x = 0,$$
$$y + V_y = 0, \quad (12)$$
$$V_z = 0.$$

The initial values of the areostationary points are chosen from (Liu et al. 2010). With these initial values, the Levenberg-Marquardt method (Levenberg 1944; Marquardt 1963; Moré 1977) based on nonlinear least-squares algorithms can be applied to solve nonlinear Eq. (12). The iteration is processed until the tolerance of Eq. (12) is less than $10^{-13}$. Finally, the positions of the areostationary points can be obtained, which are shown in Table 2.

**Table 2** Positions of areostationary points in the rotating frame.

| Areostationary Point | $x$ | $y$ | $z$ | Linear Stability* |
|---|---|---|---|---|
| $E_1$ | -0.965866684631854 | 0.259123824182275 | -0.000000206550979 | LS |
| $E_2$ | 0.259126533910926 | 0.965876790581445 | 0.000000640235545 | U |
| $E_3$ | 0.965866684631854 | -0.259123824182275 | -0.000000206550979 | LS |
| $E_4$ | -0.259126533910926 | -0.965876790581445 | 0.000000640235545 | U |

* LS = linearly stable; U = unstable.

## 5. The linear stability of areostationary points

The linear stability of areostationary points can be determined by analyzing the linearized equations of the system represented by Eq. (11) in the vicinity of areostationary points. Because of the symmetry of the system, the stabilities of areostationary points $E_1$ and $E_3$ are identical, and the stabilities of areostationary points $E_2$ and $E_4$ are identical.

Using the notation

$$\xi = x - x_e, \ \eta = y - y_e, \ \gamma = z - z_e, \tag{13}$$

and

$$\mathbf{X}(t) = (\xi, \eta, \gamma, \dot{\xi}, \dot{\eta}, \dot{\gamma})^T, \tag{14}$$

where $x_e$, $y_e$, and $z_e$ are the coordinates of the areostationary point, the linearized equations of the system represented by Eq. (11) can be expressed as

$$\dot{\mathbf{X}}(t) = \mathbf{A}(t) \cdot \mathbf{X}(t), \tag{15}$$

where

$$\mathbf{A}(t) = \begin{bmatrix} 0 & 0 & 0 & 1 & 0 & 0 \\ 0 & 0 & 0 & 0 & 1 & 0 \\ 0 & 0 & 0 & 0 & 0 & 1 \\ W_{xx} & W_{xy} & W_{xz} & 0 & 2\omega & 0 \\ W_{yx} & W_{yy} & W_{yz} & -2\omega & 0 & 0 \\ W_{zx} & W_{zy} & W_{zz} & 0 & 0 & 0 \end{bmatrix}. \qquad (16)$$

The second order partial derivatives of the effective potential $\nabla\nabla W$ must be calculated at the corresponding areostationary point.

For areostationary points $E_1$ and $E_3$, the six eigenvalues of the matrix $\mathbf{A}(t)$ are calculated as

$$\begin{aligned} \lambda_{1,2} &= \pm 0.007924462675369\,i, \quad \lambda_{3,4} = \pm 0.999892722423221\,i, \\ \lambda_{5,6} &= \pm 1.000075870390122\,i. \end{aligned} \qquad (17)$$

They are all purely imaginary, so areostationary points $E_1$ and $E_3$ are linearly stable. For areostationary points $E_2$ and $E_4$, the six eigenvalues of the matrix $\mathbf{A}(t)$ are calculated as

$$\begin{aligned} \lambda'_{1,2} &= \pm 0.007923923801517, \quad \lambda'_{3,4} = \pm 0.999945057867454\,i, \\ \lambda'_{5,6} &= \pm 1.000086331179820\,i. \end{aligned} \qquad (18)$$

Since $\lambda'_1 > 0$, areostationary points $E_2$ and $E_4$ are unstable. The states of stability of all these areostationary points appear in the fifth column of Table 2.

## 6. Periodic orbits around areostationary points

Note that around areostationary points, the displacement in the $z$ direction is much smaller than that in the $x$ or $y$ direction. The periodic orbits of the linearized system represented by Eq. (15) in the $xy$-plane can be obtained easily. Therefore, the linearized periodic solutions in the *xy*-plane can be taken as the initial guesses of periodic orbits around areostationary points in the three-dimensional space.

**6.1 Short-period orbits around linearly stable areostationary points**

The motion of the spacecraft in the *xy*-plane corresponds to the case of $C_{30} = 0$, $C_{32} = 0$, $S_{32} = 0$, and $z = 0$. Around areostationary points $E_1$ and $E_3$, the general solutions of the linearized system represented by Eq. (15) in the *xy*-plane are

$$\xi = A_1 \sin(\kappa_1 t + \varphi_1) + A_2 \sin(\kappa_2 t + \varphi_2),$$
$$\eta = \alpha A_1 \cos(\kappa_1 t + \varphi_1) + \alpha A_2 \cos(\kappa_2 t + \varphi_2), \quad (19)$$

where $\kappa_1$ and $\kappa_2$ are the frequencies; $A_1$ and $A_2$ are the amplitudes; and $\alpha = \frac{1}{2}(\kappa_2 + W_{xx}/\kappa_2)$. Since $\kappa_2 \ll \kappa_1$, the general solutions of the linearized system in the *xy*-plane consist of long-period and short-period terms.

Here, only the case of $E_1$ is considered because the cases of $E_1$ and $E_3$ are symmetrical. First, set $A_1 \neq 0$ and $A_2 = 0$; thus, only short-period terms are retained:

$$\xi = A_1 \sin(\kappa_1 t + \varphi_1), \eta = \alpha A_1 \cos(\kappa_1 t + \varphi_1). \quad (20)$$

The motion of the spacecraft is approximately periodic with the period

$$T_0 = 2\pi / \kappa_1. \quad (21)$$

For example, given amplitude $A_1 = 0.01$, the approximate periodic condition in the *xy*-plane is

$$\begin{aligned} x_0 &= -0.975525140963676, \ y_0 = 0.261715005628121, \ z_0 = 0, \\ \dot{x}_0 &= 0.005182255008665, \ \dot{y}_0 = 0.019316508183033, \ \dot{z}_0 = 0, \\ T &= 6.283859422887580. \end{aligned} \quad (22)$$

Based on the above initial guess, the Levenberg-Marquardt method is used to solve the periodic condition:

$$\mathbf{r}(t_0) - \mathbf{r}(t_0 + T) = 0,$$
$$\mathbf{v}(t_0) - \mathbf{v}(t_0 + T) = 0. \tag{23}$$

The iteration is processed until the mismatch between the position and velocity at the initial time $t_0$ and the position and velocity at the final time $t_0 + T$ is less than $10^{-12}$. The exact initial condition of the periodic orbit is derived as

$$x_0 = -0.975525140963676, \ y_0 = 0.261715005628121, \ z_0 = 0.000011183843109,$$
$$\dot{x}_0 = 0.005169425044549, \ \dot{y}_0 = 0.019268683278553, \ \dot{z}_0 = 0.000001892841807, \tag{24}$$
$$T = 6.283859507415385.$$

The periodic orbit around $E_1$ derived above is shown in Fig. 1; it has an oval shape with a period of approximately 1.03 days. In this orbit, a spacecraft can depart from the areostationary point $E_1$ by about 1.15° at most in longitude. It is expected that useful multi-view images could be derived from an offset of a few degrees, especially for observations of Elysium Planitia and Iani Chaos below the stable areostationary points $E_1$ and $E_3$, respectively.

The family of short-period orbits is shown in Fig. 2. It can be seen that the shape of the short-period orbits around $E_1$ change from oval to heart-shaped as amplitude increases.

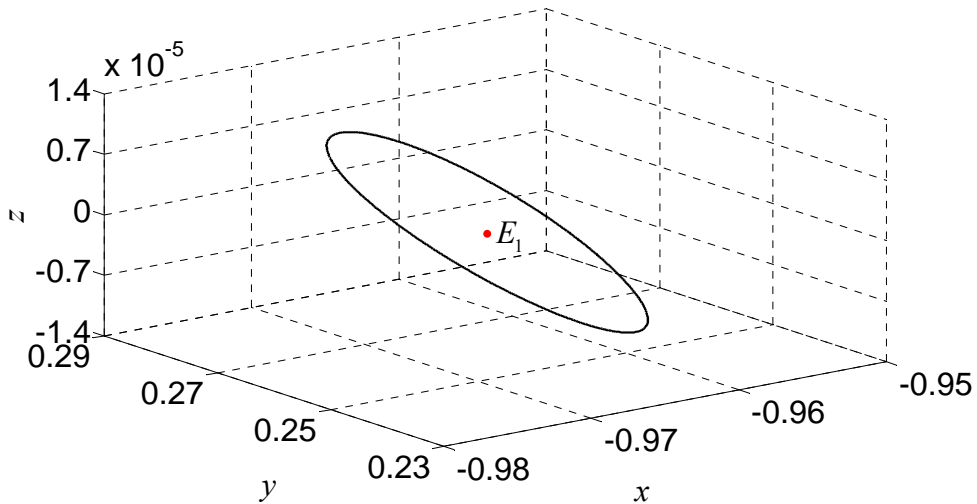

**Fig. 1** Short-period oval-shaped orbit around the areostationary point $E_1$. Different scales are used for abscissas and ordinates.

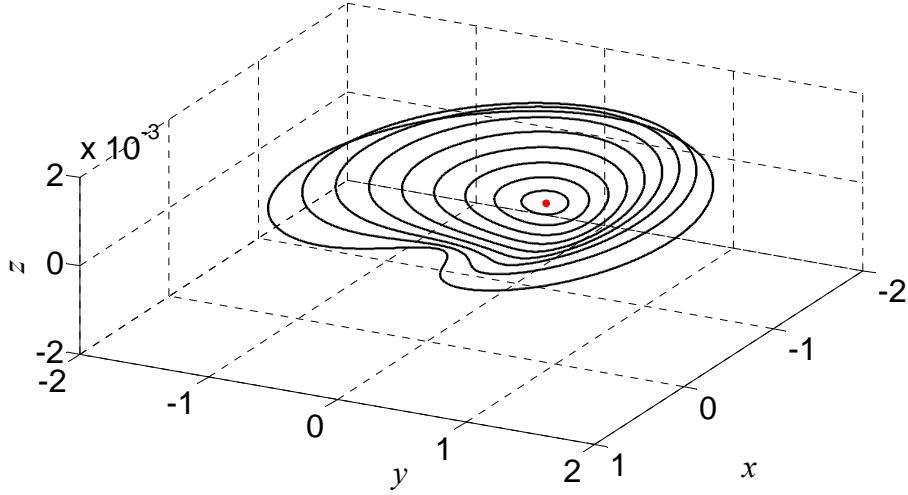

**Fig. 2** Family of short-period orbits around the areostationary point $E_1$.

According to the theory of ordinary differential equations (Robinson 2004), the fundamental solution matrix $\mathbf{\Phi}(t)$ yields the following:

$$\dot{\mathbf{\Phi}}(t) = \mathbf{A}(t)\mathbf{\Phi}(t). \tag{25}$$

The stability of periodic orbits can be determined by analyzing the eigenvalues of the monodromy matrix $\mathbf{\Phi}(T)$, where $T$ is the motion period. The monodromy matrix here is the product of the inverse of the fundamental matrix of Eq. (11) evaluated at the zero and the fundamental matrix of Eq. (11) evaluated at the period. The initial condition $\mathbf{\Phi}(0)$ used for the integration of Eq. (25) is the identity matrix. For Hamiltonian systems, the eigenvalues of $\mathbf{\Phi}(T)$ come in pairs. Each pair are reciprocal of one another. Thus, if one pair consist of two conjugate complex eigenvalues, both eigenvalues are on the unit circle, i.e., their modului are both equal to 1.

Here, the stability index $k$ is introduced to estimate the stability of the orbit, and is defined as the sum of all the eigenvalues of $\Phi(T)$:

$$k = \sum_{i=1}^{6} |\chi_i|, \qquad (26)$$

where $\chi_i$ is the eigenvalue of $\Phi(T)$. Therefore, if $k > 6$, the orbit is unstable; if $k = 6$, the orbit is stable.

The general behavior of the stability index $k$ as a function of the Jacobian constant is presented in Fig. 3. It can be seen that the $E_1$ family of periodic orbits changes from stable ($k = 6$) to unstable ($k > 6$) as the value of the Jacobian constant increases.

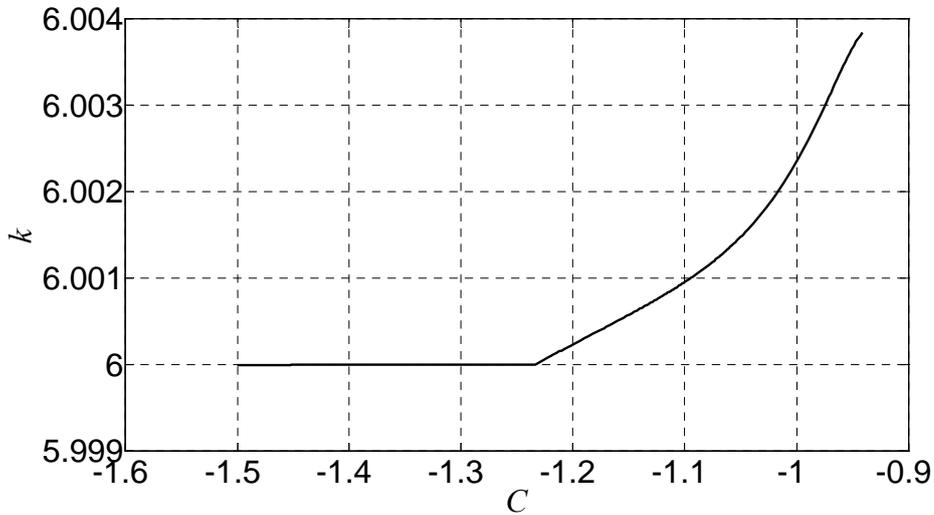

**Fig. 3** Evolution of stability index $k$ as a function of the Jacobian constant $C$.

**6.2 Long-period orbits around linearly stable areostationary points**

When $A_1 = 0$ and $A_2 \neq 0$, only long-period terms are retained:

$$\xi = A_2 \sin(\kappa_2 t + \varphi_2), \eta = \alpha A_2 \cos(\kappa_2 t + \varphi_2). \tag{27}$$

If $A_1 = 0$ and $A_2 = 0.001$, Eq. (27) is taken as the initial guess, and the Levenberg-Marquardt method is utilized until the mismatch between the position and velocity at the initial time $t_0$ and the position and velocity at the final time $t_0 + T$ is less than $10^{-13}$. Then, the exact condition of the periodic orbit can be obtained as

$$\begin{aligned}&x_0 = -0.966832530336112,\ y_0 = 0.259382942061755,\ z_0 = -0.000000206142371,\\ &\dot{x}_0 = 0.000388605225698,\ \dot{y}_0 = 0.001448499930975,\ \dot{z}_0 = 0.000000000052347, \\ &T = 8.001262797809567e+002.\end{aligned} \tag{28}$$

The long-period orbit around the areostationary point $E_1$ is shown in Fig. 4. The period of the orbit is approximately 0.36 years. For this long-period orbit, the six eigenvalues of the monodromy matrix $\mathbf{\Phi}(T)$ are all on the unit circle, and it is easy to see that $k = 6$ based on Eq. (26). Thus, this long-period orbit around the areostationary point $E_1$ is stable.

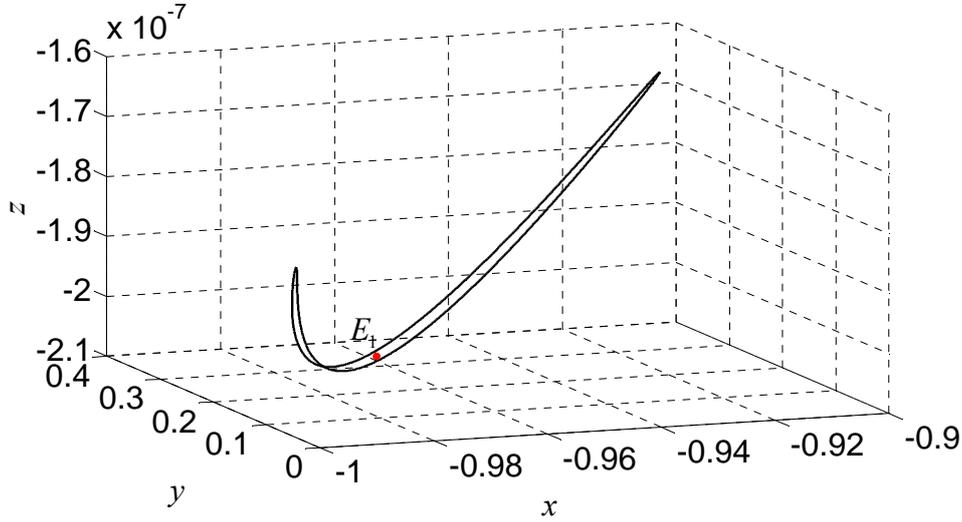

**Fig. 4** Long-period orbit around the areostationary point $E_1$.

### 6.3 Short-period orbits around unstable areostationary points

Around areostationary points $E_2$ and $E_4$, the general solutions of the linearized system represented by Eq. (15) in the $xy$-plane are

$$\xi = D_1 e^{\kappa_1' t} + D_2 e^{-\kappa_1' t} + A' \sin(\kappa_2' t + \varphi'),$$
$$\eta = \alpha_1' \left(D_1 e^{\kappa_1' t} - D_2 e^{-\kappa_1' t}\right) + \alpha_2' A' \cos(\kappa_2' t + \varphi'), \quad (29)$$

where $\kappa_1'$ and $\kappa_2'$ are the frequencies; $D_1$, $D_2$, and $A'$ are the amplitudes; and $\alpha_1' = \frac{1}{2}(\kappa_1' - W_{xx}/\kappa_1')$ and $\alpha_2' = \frac{1}{2}(\kappa_2' + W_{xx}/\kappa_2')$. It can be seen that Eq. (29) consists of secular terms, attenuation terms, and periodic terms. In order to derive periodic orbits, the coefficient of secular term $D_1$ and the coefficient of attenuation term $D_2$ must be set equal to zero, so

$$\xi = A' \sin(\kappa_2' t + \varphi'), \eta = \alpha_2' A' \cos(\kappa_2' t + \varphi'). \quad (30)$$

Based on the Levenberg-Marquardt method, the initial guess of the periodic orbit represented by Eq. (30) can be adjusted slightly. The iteration is processed until the mismatch between the position and velocity at the initial time $t_0$ and the position and velocity at the final time $t_0 + T$ is less than $10^{-12}$. Taking the case of $A' = 0.01$ as an example, the exact initial condition of the periodic orbit can be obtained as

$$x_0 = 0.261717716762586, \ y_0 = 0.975535246535669, \ z_0 = -0.000045126967204,$$
$$\dot{x}_0 = 0.019269493269170, \ \dot{y}_0 = 0.005169641770384, \ \dot{z}_0 = -0.000002453663158, (31)$$
$$T = 6.283530658585495.$$

An oval-shaped periodic orbit around the areostationary point $E_2$ is shown in Fig. 5; its period is approximately 1.03 days. Fig. 6 shows that the shapes of the periodic orbits around $E_2$ also change from oval to heart-shaped as amplitude increases. Fig. 7 presents the evolution of the stability index $k$ as a function of the Jacobian constant. It is clear that the $E_2$ family of periodic orbits changes from unstable ($k > 6$) to stable ($k = 6$) as the value of the Jacobian constant increases. It can be seen that stable periodic orbits exist around unstable areostationary points. This means that a spacecraft could stay in the vicinity of an unstable areostationary point for a long time, which would facilitate observation of Martian topography, including Syrtis Major Planum and Tharsis below unstable areostationary points $E_2$ and $E_4$, respectively.

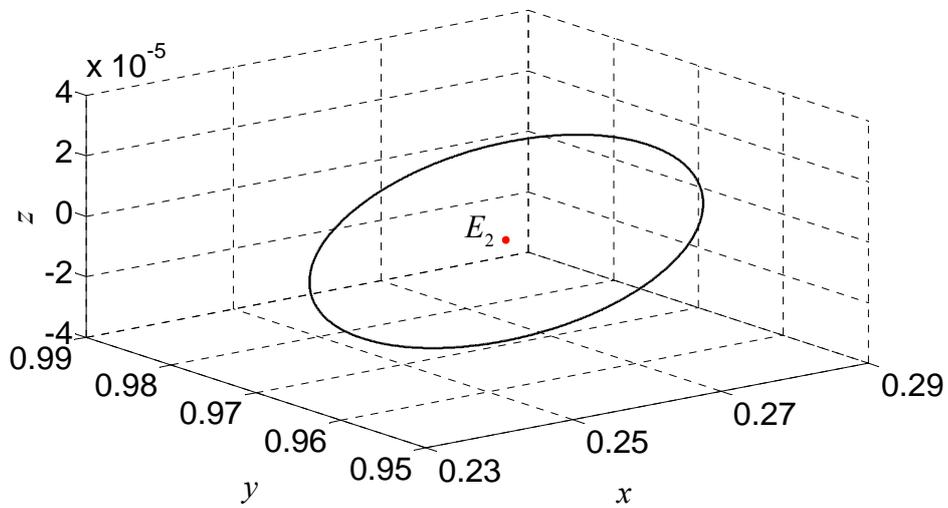

**Fig. 5** Oval-shaped periodic orbit around the unstable areostationary point $E_2$.

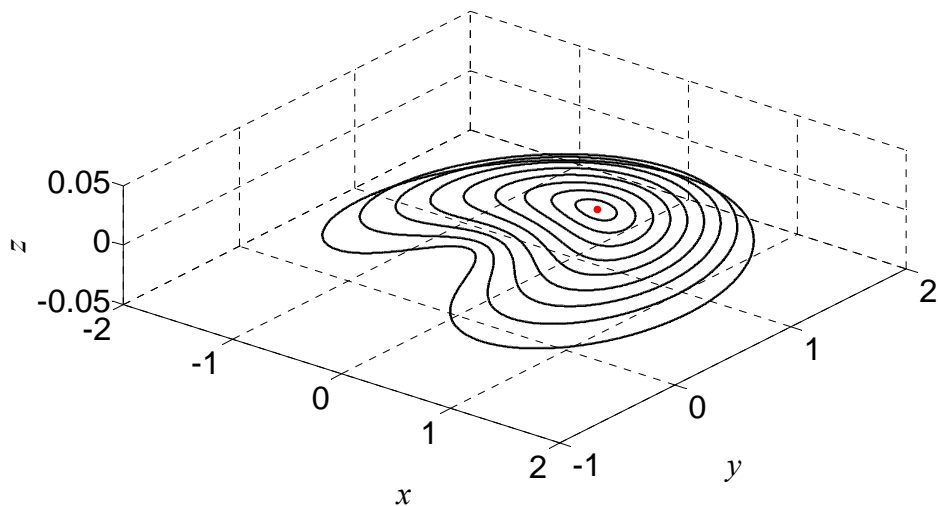

**Fig. 6** Family of short-period orbits around the areostationary point $E_2$.

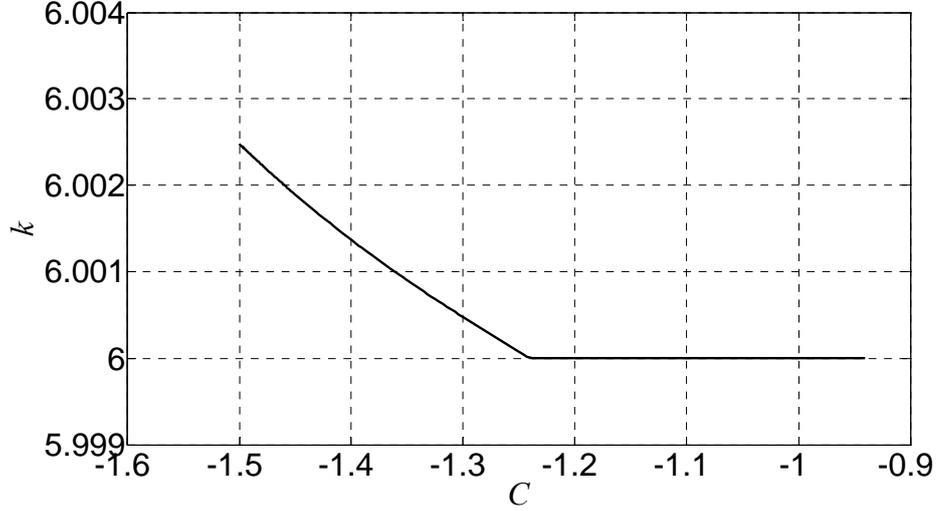

**Fig. 7** Evolution of the stability index $k$ as a function of the Jacobian constant $C$.

## 6.4 Vertical period orbits around both linearly stable and unstable areostationary points

Vertical periodic orbits around both linearly stable and unstable areostationary points are also noted, which are different from both short-period and long-period orbits. The motion of the spacecraft in the $z$ direction corresponds to the case of $\xi = 0$, and $\eta = 0$.

Around all areostationary points, it is easy to obtain that the eigenvalues of Eq. (15) in the $z$ direction are all purely imaginary. Thus, the general solutions of the linearized system represented by Eq. (15) in the $z$ direction are calculated as

$$\gamma = B\sin(\varsigma t + \theta), \tag{32}$$

where $\varsigma$ is the frequency, and $B$ is the amplitude. Based on the initial guess represented by Eq. (32), the periodic condition (23) can be solved using the Levenberg-Marquardt method. The iteration is processed until the mismatch between the position and velocity at the initial time $t_0$ and the position and velocity at the final time $t_0 + T$ is less than $10^{-12}$.

For example, given the amplitude $A = 10^{-5}$ around the areostationary point $E_1$, the exact initial condition of the vertical periodic orbit can be obtained as

$$
\begin{aligned}
&x_0 = -0.965866684631854, \quad y_0 = 0.259123824182275, \quad z_0 = 0.000008038019607,\\
&\dot{x}_0 = 0, \quad\quad\quad\quad\quad\quad\quad\quad\; \dot{y}_0 = 0, \quad\quad\quad\quad\quad\quad\quad\;\; \dot{z}_0 = 0.000000506705147, \quad (33)\\
&T = 6.282708621687184.
\end{aligned}
$$

The vertical periodic orbit around $E_1$ derived above is shown in Fig. 8; its period is approximately 1.03 days. For this vertical periodic orbit, the six eigenvalues of the monodromy matrix $\boldsymbol{\Phi}(T)$ are all on the unit circle, and it is easy to see that $k = 6$ based on Eq. (26). Thus, this vertical periodic orbit around the linearly stable areostationary point $E_1$ is stable.

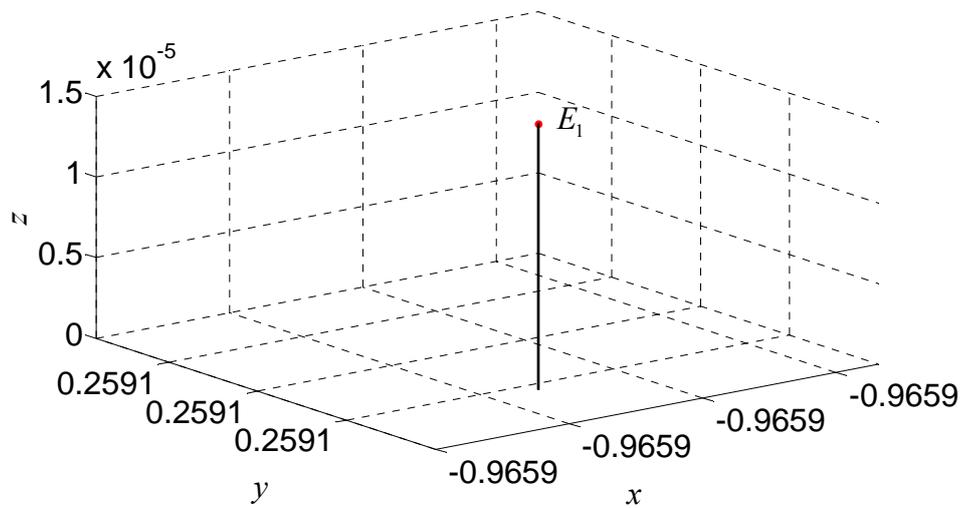

**Fig. 8** Vertical periodic orbit around the linearly stable areostationary point $E_1$.

Given amplitude $A = 10^{-5}$ around the areostationary point $E_2$, the exact initial condition of the vertical periodic orbit can be obtained as

$x_0 = 0.259126533910926,\quad y_0 = 0.965876790581445,\quad z_0 = 0.000008320016725,$
$\dot{x}_0 = 0,\qquad\qquad\qquad\quad \dot{y}_0 = 0,\qquad\qquad\qquad\quad \dot{z}_0 = 0.000001413345788,$ (34)
$T = 6.282642913717483.$

The vertical periodic orbit around $E_2$ derived above is shown in Fig. 9; its period is also approximately 1.03 days. For this vertical periodic orbit, it is easy to obtain that $k = 6.0025$ based on Eq. (26). Thus, this vertical periodic orbit around the unstable areostationary point $E_2$ is unstable. Note that the vertical periodic orbits around areostationary points $E_1$ and $E_2$ are both below the areostationary points.

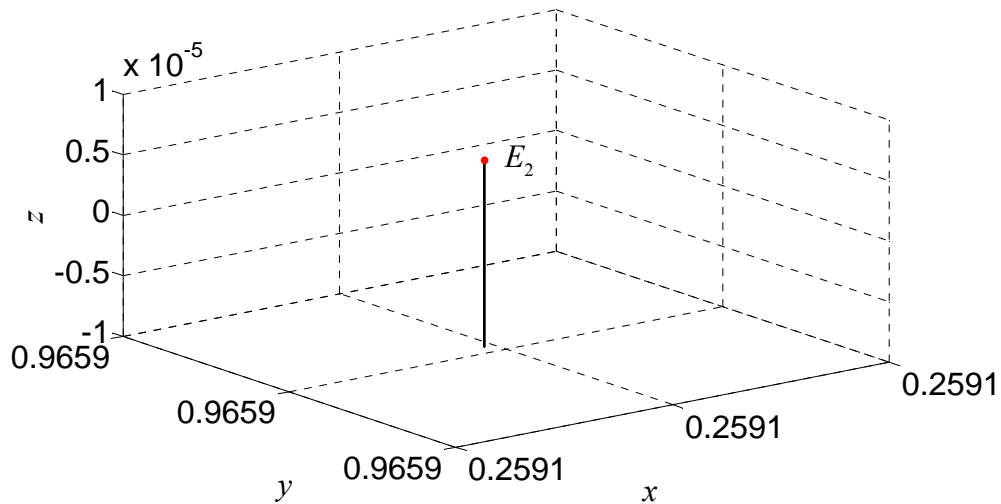

**Fig. 9** Vertical periodic orbit around the unstable areostationary point $E_2$.

# 7. Heteroclinic orbits connecting the two unstable areostationary points

The numerical method used to calculate the stable and unstable manifolds in this study was developed in previous research (Robinson 2004). The local stable and unstable manifolds of areostationary points are approximated with line segments through the areostationary points in the directions of stable eigenvector $v^s$ and unstable eigenvector $v^u$; these eigenvectors can be obtained easily by analyzing the linearized equation in the vicinity of the areostationary point.

Take the calculation of the unstable manifold of the areostationary point $E_2$ as an example. For a sufficiently small $\delta$ (given as $10^{-9}$ here), a point $X$ in the neighborhood of the areostationary point $E_2$ is selected so that

$$X = E_2 + \delta v^u. \qquad (35)$$

Taking $X$ as the initial value, the global unstable manifold can be obtained by integrating Eq. (11) forward. Following a similar procedure, the global stable manifold can also be obtained by backward iteration.

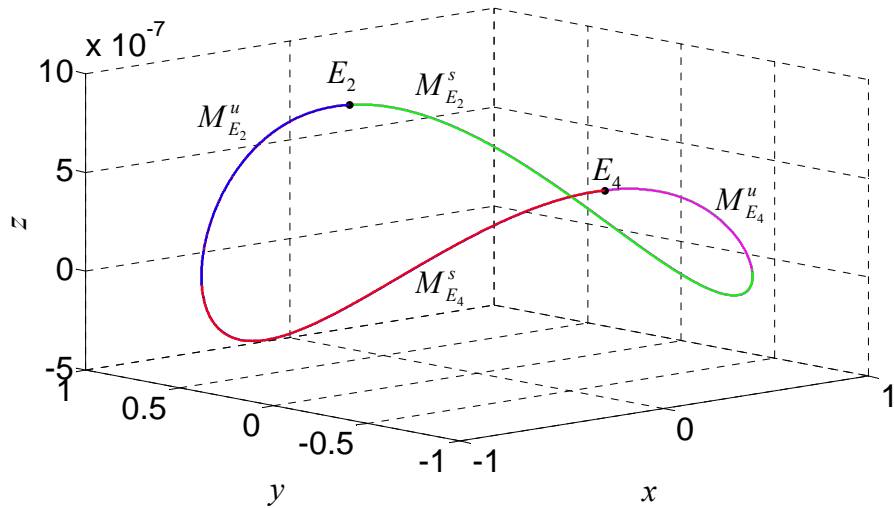

**Fig. 10** Heteroclinic orbits connecting unstable areostationary points $E_2$ and $E_4$. Line in red corresponds to the unstable manifold of $E_2$; line in blue corresponds to the stable manifold of $E_4$; line in pink corresponds to the unstable manifold of $E_4$; and line in green corresponds to the stable manifold of $E_2$.

Fig. 10 illustrates the stable and unstable manifolds of areostationary points $E_2$ and $E_4$. Note that the unstable manifold $M_{E_2}^u$ of $E_2$ coincides with the stable manifold $M_{E_4}^s$ of $E_4$, which is known as the heteroclinic orbit connecting unstable areostationary points $E_2$ and $E_4$. It can also be seen that the unstable manifold $M_{E_4}^u$ of $E_4$ coincides with stable manifold $M_{E_2}^s$ of $E_2$, which is another heteroclinic orbit connecting $E_2$ and $E_4$. The existence of heteroclinic orbits means that a spacecraft at the unstable areostationary point could transfer to another unstable areostationary point with minimal energy consumption.

## 8. Conclusions

This study analyzes areostationary orbits around Mars in the three-dimensional space. Areostationary points are derived in the rotating reference frame, and it is shown that periodic orbits exist around both linearly stable and unstable areostationary points. For short-period orbits around linearly stable areostationary points, their stabilities evolve from stable to unstable as the Jacobian constant increases. For short-period orbits around unstable areostationary points, their stabilities evolve from unstable to stable as the Jacobian constant increases. It is also determined that vertical periodic orbits around linearly stable areostationary points are stable, while vertical periodic orbits around unstable areostationary points are unstable. Finally, heteroclinic orbits connecting the two unstable areostationary points are found, providing the possibility for orbital transfer with minimal fuel expenditure.

## Acknowledgments

This work was supported by National Basic Research Program of China (973 Program) (2012CB720000) and the National Natural Science Foundation of China (No. 11072122).

## References

Alvarellos, J. L. 2011, The Journal of the Astronautical Sciences, 57

Anderson, R., Bar-Sever, Y., Bell, D., Ely, T., Guinn, J., Hart, M., Kallemeyn, P., Levene, E., Jah, M., Romans, L., & Wu, S. 1999, in International Symposium on Space Communications and Navigation Technologies (Pasadena, California)

Bell, D. J., Cesarone, R. J., Ely, T. A., Edwards, C. D., & Townes, S. A. 2000, in IEEE Aerospace Conference proceedings, Vol. 7 (Big Sky, MT), 75

Cesarone, R., Hastrup, R., Bell, D., Lyons, D., & Nelson K. 1999, in AAS/AIAA Astrodynamics Specialist Conference (Girdwood, Alaska)

Clarke, A. C. 1945, Wireless World, 11, 305

Deprit, A., & López, T. 1996, Revista Matemática Universidad Complutense, 9, 311

Edwards, C. D., Adams, J. T., Agre, J. R., Bell, D. J., Clare, L. P., Durning, J. F., Ely, T. A., Hemmati, H., Leung, R. Y., & McGraw, C. A. 2000, in Concepts and Approaches for Mars Exploration (Houston), 105

Edwards, C. D. & Depaula, R. 2007, Acta Astronautica, 61, 131

Edwards, C. D. 2007, International Journal of Satellite Communications and Networking, 25, 111

Hastrup, R. C., Bell, D. J., Cesarone, R. J., Edwards, C. D., Ely, T. A., Guinn, J. R., Rosell, S. N., Srinivasan, J. M., & Townes, S. A. 2003, Acta Astronautica, 52, 227

Hu, W., & Scheeres, D. J. 2008, RAA (Chinese Journal of Astronomy and Astrophysics), 8, 108

Kaula, W. M., 1966, Theory of satellite geodesy: applications of satellites to geodesy (Waltham, Blaisdell)

Konopliv, A. S., Asmar, S. W., Folkner, W. M., Karatekin, Ö., Nunes, D. C., Smrekar, S. E., Yoder, C. F., & Zuber, M. T. 2011, Icarus, 211, 401

Lara, M., & Elipe, A. 2002, Celestial Mechanics and Dynamical Astronomy, 82, 285

Levenberg, K. 1944, Quarterly Applied Mathematics, 2, 164

Liu, X., Baoyin, H., & Ma, X. 2010, Journal of Guidance, Control, and Dynamics, 33, 1294

Marquardt, D. W. 1963, SIAM Journal Applied Mathematics, 11, 431

Moré, J. J. 1977, in Lecture Notes in Mathematics 630: Numerical Analysis, ed., G. A. Watson (Dundee; Springer Verlag), 105

Robinson, R. C., 2004, An introduction to dynamical systems: continuous and discrete (Upper

Saddle River, Pearson Prentice Hall)

Scheeres, D. J., Durda, D. D., & Geissler. P. E. 2002, in Asteroids III, eds., W. F. Bottke et al. (Tucson; University of Arizona Press), 527

Turner, A. E. 2006, in 24th AIAA International Communications Satellite Systems Conference, AIAA-2006-5313 (San Diego, California)

Thomas P. C. 1993, Icarus, 105, 326

Wytrzyszczak, I. 1998, Artificial Satellites, 33, 11

Zhuravlev, S. G. 1977, Astronomicheskii Zhurnal (Astronomy Reports), 54, 909